\documentclass[10pt,a4paper]{article}

\usepackage{epsfig}
\usepackage{subfigure}
\usepackage{calc}
\usepackage{amssymb}
\usepackage{amstext}
\usepackage{amsmath}
\usepackage{amsthm}
\usepackage{multicol}
\usepackage{pslatex}
\usepackage{apalike}
\usepackage{enumitem} 

\graphicspath{{./figs/}}

\usepackage{array}
\newcolumntype{P}[1]{>{\centering\arraybackslash}p{#1}}

\usepackage{url}

\title{Evolution and Perspectives of the Keep IT Secure Ecosystem:
A Six-Year Analysis of Cybersecurity Experts Supporting Belgian SMEs }

\author{Christophe Ponsard\textsuperscript{1}, Jean-François Daune\textsuperscript{1}, Denis Darquennes\textsuperscript{1},\\
Malik Bouhou\textsuperscript{1}, and Nicolas Point\textsuperscript{2}\\
\textsuperscript{1}CETIC Research Centre, Charleroi, Belgium\\
\textsuperscript{2}Multitel Research Centre, Mons, Belgium\\
\texttt{\{cp,jfd,dda,mbo\}@cetic.be, point@multitel.be}
}

\date{January 2026}

\begin{document}

\maketitle



\begin{abstract}
The importance of cybersecurity for Small and Medium Enterprises (SMEs) has never been greater, especially given the rise of AI-driven threats. Supporting SMEs requires a sustained effort to ensure they have access to resources and expertise covering awareness, protection, auditing, and incident response. Since 2019, our work with the Keep It Secure initiative has focused on helping Belgian (Walloon) SMEs strengthen their cybersecurity posture through access to a network of labelled cybersecurity experts. In this process, we interviewed over 120 professionals from around 90 companies and gathered rich insights about the nature, strengths and weaknesses of our regional ecosystem. While our initiative primarily targets the labelling of cybersecurity experts, we demonstrate increasing alignment with the broader Cyber Fundamentals framework deployed at the federal level in Belgium, which supports official certification. This paper reports on the progress and lessons learned from this long-term effort, highlighting how expert validation, based on a structured evaluation approach, can help improve SME cybersecurity.
\end{abstract}


\section{Introduction}

Small and Medium Enterprises (SMEs) are a major driver of socio-economic development, contributing more than half of Europe’s economic value and employing roughly two-thirds of the workforce \cite{EU14}. As digitalisation accelerates, IT systems have become business-critical, yet many SMEs lack the resources, expertise, or time to fully protect themselves against cybersecurity threats. Many continue to underestimate their exposure or assume their size makes them unattractive targets \cite{Alahmari2020}. 

In reality, SMEs are increasingly targeted by cybercriminals. The threat landscape has evolved rapidly in recent years. AI-driven attacks deploying automated phishing, adaptive malware, reconnaissance tools, and deepfakes have amplified the scale, speed, and sophistication of attacks \cite{Adlercreutz25}. Recurring studies estimate that a majority of SMEs have experienced at least one significant cyber incident, with more than half of them failing to recover and ceasing operations within months \cite{NCSA-Guide}. Moreover, while awareness about cybercrime is rising, this does not translate into effective preparedness or efficient risk management, highlighting the need for practical cybersecurity strategies tailored to SME constraints \cite{Arroyabe24}. This is widely acknowledged and supported at European level by organisations such as ENISA, SME Alliance, the European Commission, and European Cybersecurity Organisation (ECSO). At national level, most countries have set up some form of programme to raise awareness and to provide guidance as reported in our previous work \cite{Ponsard18,Ponsard19}. Examples of such initiatives are the CyberEssentials in the UK \cite{cyberessentials} or the Finnish Cyber Security Certificate \cite{FINCSC}. 

\begin{figure}[h]
\centering
\includegraphics[width=0.95\textwidth]{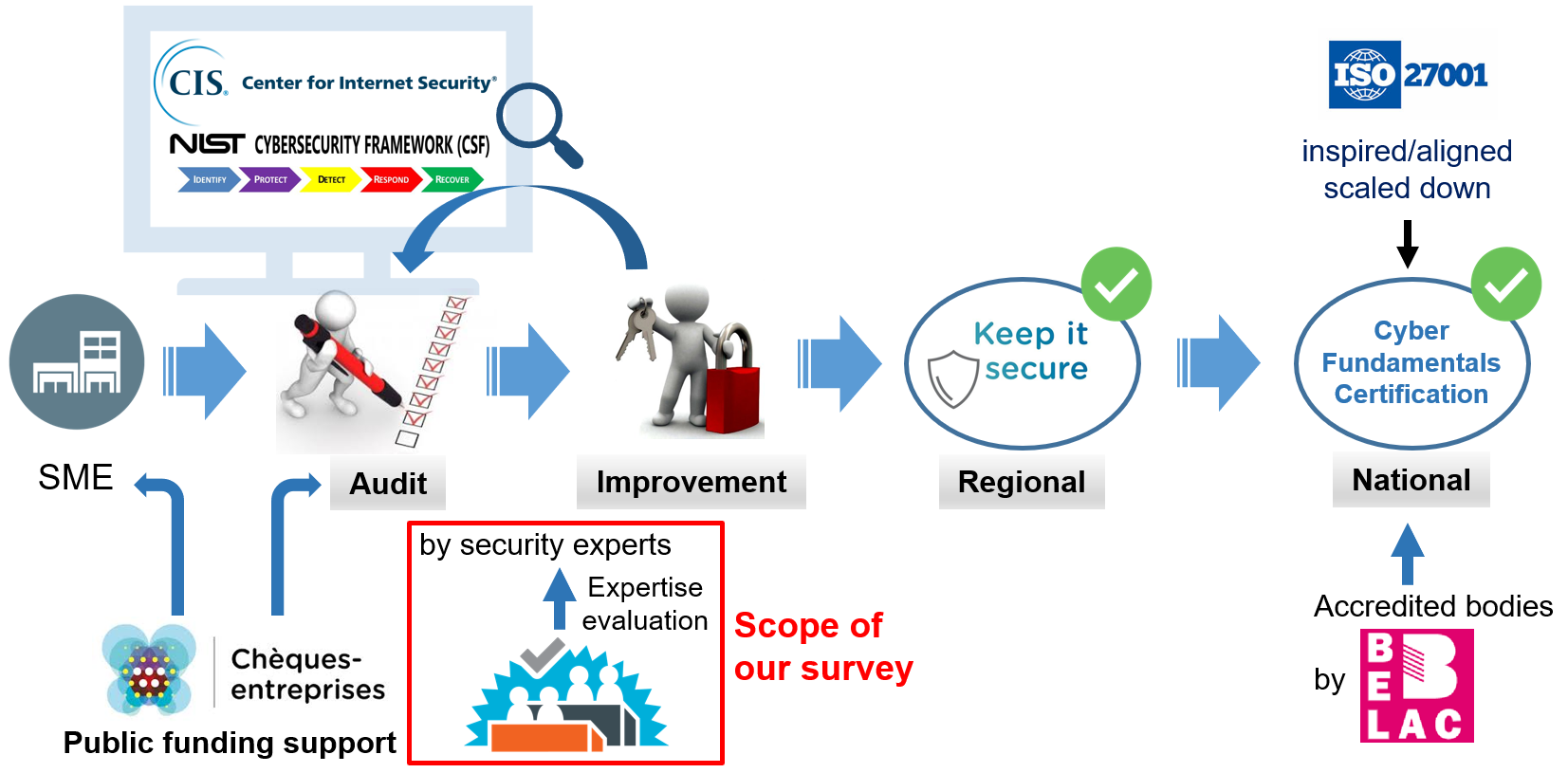}
\caption{KIS Ecosystem in Belgium and scope of this survey (source: \cite{Ponsard20b})}
\label{fig:KIS}
\end{figure}

In Belgium, the effort is currently structured at the two levels depicted in Figure \ref{fig:KIS}: 
\begin{itemize}
    \item \textit{the regional level (Wallonia)} is issuing labels for cybersecurity experts through non-certifying audits with the aim to make sure SMEs have access to skilled people to help them identify risks and deploying adequate measures to manage them. Our initiative called Keep IT Secure (KIS for short) went in operation in 2019. Cybersecurity expertise and ability to advise SMEs are validated by an advisory committee. Jointly, a funding scheme is available to support the intervention of labelled experts.
    \item \textit{the national level} is concerned about providing a certification scheme called Cyber Fundamentals (CyFun for short); inspired by NIST CSF \cite{NIST-CSF} and lighter than ISO27K \cite{ISO27K}. This scheme targets all companies and proposes different profiles adapted to the threat level: basic, important and essential. In the scope of this paper, we will focus on the basic level which is adapted to standard SMEs \cite{CyFun25}.
\end{itemize}

The purpose of this paper is to report about the evolution of our regional ecosystem of cybersecurity experts based on a solid dataset of more than 120 evaluation questionnaires collected over a period of six years with a well-defined and stable methodology based on the NIST CSF framework and the CIS controls. Our research question can be expressed as: ``To what extent does the Keep IT Secure lightweight labelling framework improve the maturity of a regional cybersecurity expertise ecosystem over time, and what key drivers of improvement can be identified?'' The methodology is recapped to highlight how it was conducted during the period without breaking continuity while taking into account evolutions like the convergence with the federal CyFun framework. We analyse the evolution in size and maturity and identify some lessons learned related to  main weaknesses, new skills to consider (e.g. CTI, AI threats) and the regulation driver (GDPR, NIS2, AI Act).

Our work is structured as follows. First, Section 2 recalls about our evaluation methodology presented in \cite{Ponsard19}. We compare it to CyFun 2025 to show the current level of convergence. Section 3 reports on our application of the resulting expert validation toolkit over a six-years period with a state of the art AI-powered analytic framework. Then Section 4 present the main lessons we learned and some planned evolution. Finally, Section 5 draws some conclusions and sketches our future work.

\section{Background on KIS}

\subsection{KIS Audit Process}

Keep IT Secure goal is part of a larger mechanism to support Walloon SME in their innovation and digitalisation efforts by partly supporting the funded intervention of dedicated experts, in our case in cybersecurity. Unlike other domains where ex-post verification is possible, the sensitive nature of cybersecurity requires making sure that service providers are qualified experts prior to their intervention. This includes their ability to identify cybersecurity risks that may impact the SME business and suggest adequate protection mechanisms. 

KIS checks that an expert (\underline{and not the company}): 
\begin{itemize}[noitemsep, topsep=0pt]
\item knows about key cybersecurity concepts and reference frameworks.
\item understands of SME-specific cybersecurity issues.
\item can carry out a technical audit, using a well-established methodology that may be their own.
\end{itemize}

KIS was designed around 2018. At that time we reviewed the few available frameworks targetting SMEs \cite{Ponsard18}. We came up with a method to interview experts and to check the above requirements without assuming they follow an imposed methodology. Instead, we evaluate the coverage of fundamental practices that ensure a sound level of cybersecurity maturity within SMEs based on a concrete scenario used as support for a dynamic discussion to evaluate the following aspects:
\begin{itemize}[noitemsep, topsep=0pt]
\item identification of risks w.r.t SME context.
\item main strategies using NIST CSF1.1 \cite{NIST-CSF} covering identification, protection, detection, response, and recovery.
\item use of controls, based on a check-list inspired by Center for Internet Security \cite{CIS16} with a focus on basic controls but also inclusion of some intermediate and more advanced controls with a lesser weight on the evaluation.
\end{itemize}

\begin{figure}[h]
\centering
\includegraphics[width=\columnwidth]{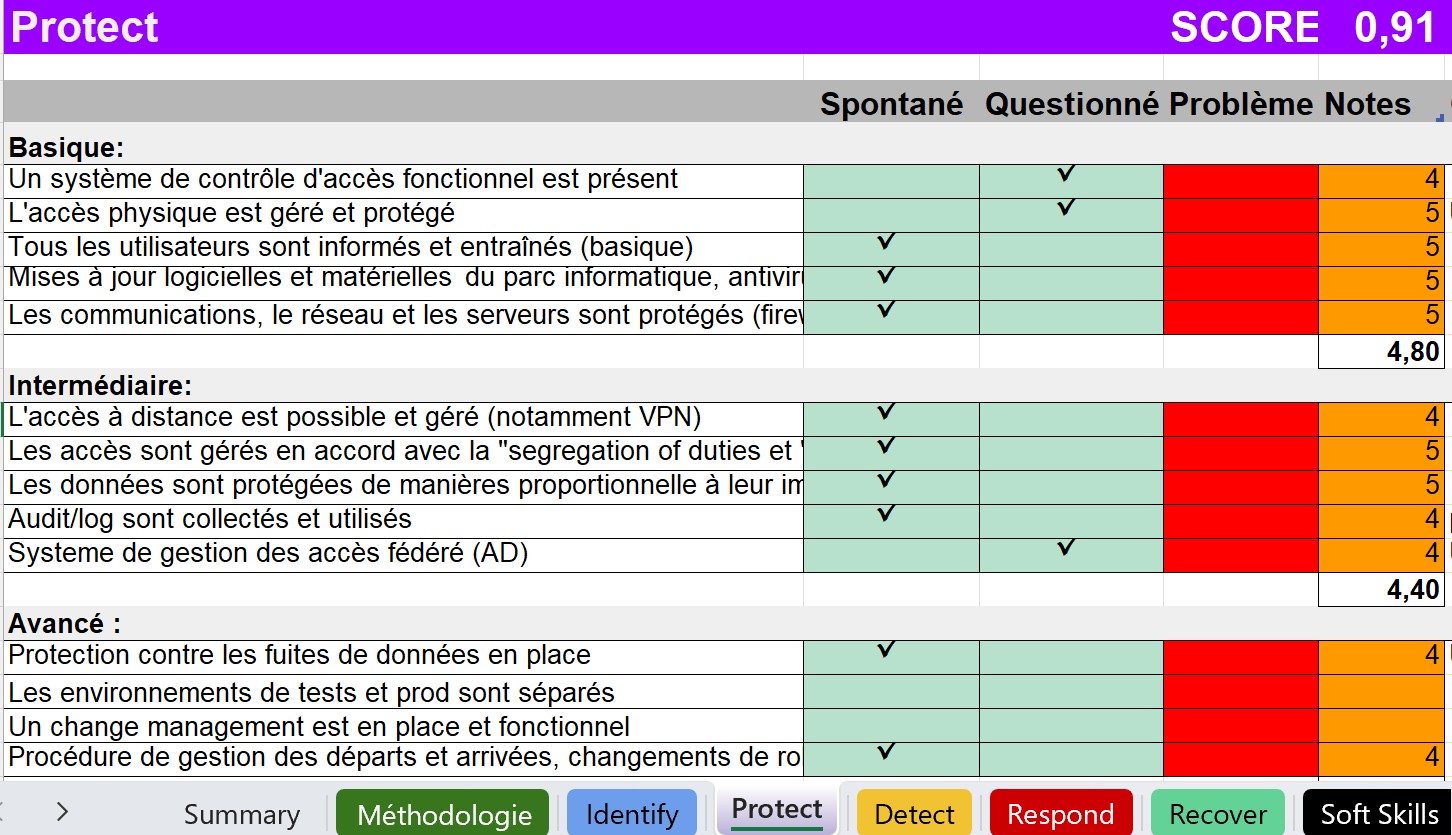}
\caption{Identify tab of the KIS assessment form}
\label{fig:KIS-form}
\end{figure}

The interview is led by two specialists from the cybersecurity advice centre and lasts for a maximum of two hours. After welcoming the candidate, he is first asked to describe his professional training and experience. Then a case is presented (e.g. a grocery store using IT for inventory, billing, procurement and a basic website). The candidate is asked to report how it deals with the cybersecurity audit. His method is evaluated against our checklist of about 50 controls. We record in a spreadsheet if proposed actions were reported spontaneously, after giving some hint or were omitted or wrong. The sheet has several tabs to covers all the NIST CSF steps and some extra tabs e.g. for summary. An example of filled form tab for Identify phase is shown in Figure \ref{fig:KIS-form}. 

Our approach revealed a great choice in terms of stability because the next CIS version aligned with NIST CSF and our the federal CyFUn framework developed around 2021 adopted a similar approach (partly based on our feedback). As a result, the method could be used with minor adjustments over the six year period reported here. The operation was actually divided into 3 different funding periods with the following evolution:
\begin{itemize}[noitemsep, topsep=0pt]
\item \textbf{2019-2021}: initial questionnaire recording only key characteristics of answers (spontaneous, asked, basic, wrong) and a comment. Those are turned into scores using a weighted formula with respectively weights 3,2,1,0. Questions are also weighted depending on their type: basic (50\%), intermediate (35\%) and advanced (15\%).
\item \textbf{2021-2023}: quantitative score (0-5) enabling to take answer quality into account. For example, a spontaneous incomplete answer could be ranked less than a perfect answer given after an hint. Explicit sheet for soft skills was also introduced.
\item \textbf{2024-now}: minor update, especially about presentation using radar charts.
\end{itemize}



\subsection{CyFun Alignment}

Although their goals differ, it makes sense to have KIS and CyFun aligned. We said there is no imposed method but CyFun basic is a recommended one because it is the baseline to consider for an SME and it opens the way to certifying at that level. Consequently, it also makes sense to consider it as a reference when auditing a candidate KIS expert for labelling. Table \ref{tab:comparison} compares the latest version of KIS and CyFun. It shows a very good alignment on most of the categories. This is not surprising given the common grounds and the co-evolution mentioned earlier. The current differences are:
\begin{itemize}[noitemsep]
\item CyFun has adopted CSF2.0 with a Govern section. Those topics are present in Identify in KIS. CyFUn is a bit more detailed regarding legal issues and human resources.
\item Risk focus more on nature and level in KIS while CyFun considers threats and vulnerabilities.
\item Protect is very similar with KIS focusing only on updates. Backups are managed here for CyFun while they are analysed in Recovery for KIS.
\item Detect considers responsibility in KIS and correlation in CyFun.
\item Respond and recover use a different terminology for the involved plans.
\end{itemize}

\begin{table}[h]
\small
\begin{center}
\begin{tabular}{|>{\raggedright\arraybackslash}p{2cm}|>{\raggedright\arraybackslash}p{3.5cm}|>{\raggedright\arraybackslash}p{3.5cm}|}
\hline
\textbf{Topic} & \textbf{KIS} & \textbf{CyFun} \\
\hline
Origin & 2019 & \textasciitilde 2021 (draft) \\
\hline
Focus & RW & Federal \\
\hline
Purpose & Contractor labelling & Certification \\
\hline
Target & SMEs & Any company \\
\hline
Revisions & 2022, 2024 & 2023, 2025 \\
\hline
Basis & NIST CSF 1.1, CIS20 & NIST CSF2 (v2025) \\
\hline
Levels & Basic (50\%) \newline Intermediate (85\%) \newline Advanced (100\%) & (Small) Basic \newline Important \newline Essential \\
\hline
Evaluation & Expert eval. form & Self-assessment \newline External audit \newline Govern (CSF2.0) \\
\hline
Identify \newline (Govern)& Risk process \newline Security policy \newline Asset inventory \newline Risk (nature, level) & Risk management \newline Policy + legal + HR\textsuperscript{*} \newline Asset inventory \newline Risks (threats/vuln.) \newline Improvements \\
\hline
Protect & Access control, Physical \newline Data, logs\newline Network \newline SW\textsuperscript{*} updates \newline Information/training & Access control, Physical access \newline Data, backups, logs\newline Network \newline SW instal/execution \newline Awareness \\
\hline
Detect & Monitoring (generic) \newline Responsibility & Monitoring (network/staff) \newline \newline Correlation \\
\hline
Respond & IRP\textsuperscript{*} (BCP part\textsuperscript{*}) \newline Notification & IRP\textsuperscript{a} \newline Notifications \\
\hline
Recover & IRP\textsuperscript{*} (DRP\textsuperscript{*} part) \textsuperscript{a}\newline Communication \newline Backup & IRP\textsuperscript{*} (DRP\textsuperscript{*} part) \\
\hline
\end{tabular}
\end{center}
\vspace{-2mm} 
\small{
\textsuperscript{*} SW = Software, HR =  Human Resources, IRP = Incident response plan,\\ 
\textsuperscript{~~} BCP = Business Continuity Plan, DRP = Disaster Recovery Plan
}
\caption{Comparison of KIS3 and CyFun 2025}
\label{tab:comparison}
\end{table}

\newpage
\section{Analysis}

This section reports on our analysis of 122 evaluations between March 2019 and June 2025. After presenting the analysis process, anonymised aggregated statistics are analysed across the following dimensions: company-level characterisation (business sector, size, areas of expertise), statistics about acceptance rate and average score, detailed maturity assessment, and global evolution of KIS ecosystem.

\subsection{Data Analysis Process}

\begin{figure}[h]
\centering
\includegraphics[width=0.55\columnwidth]{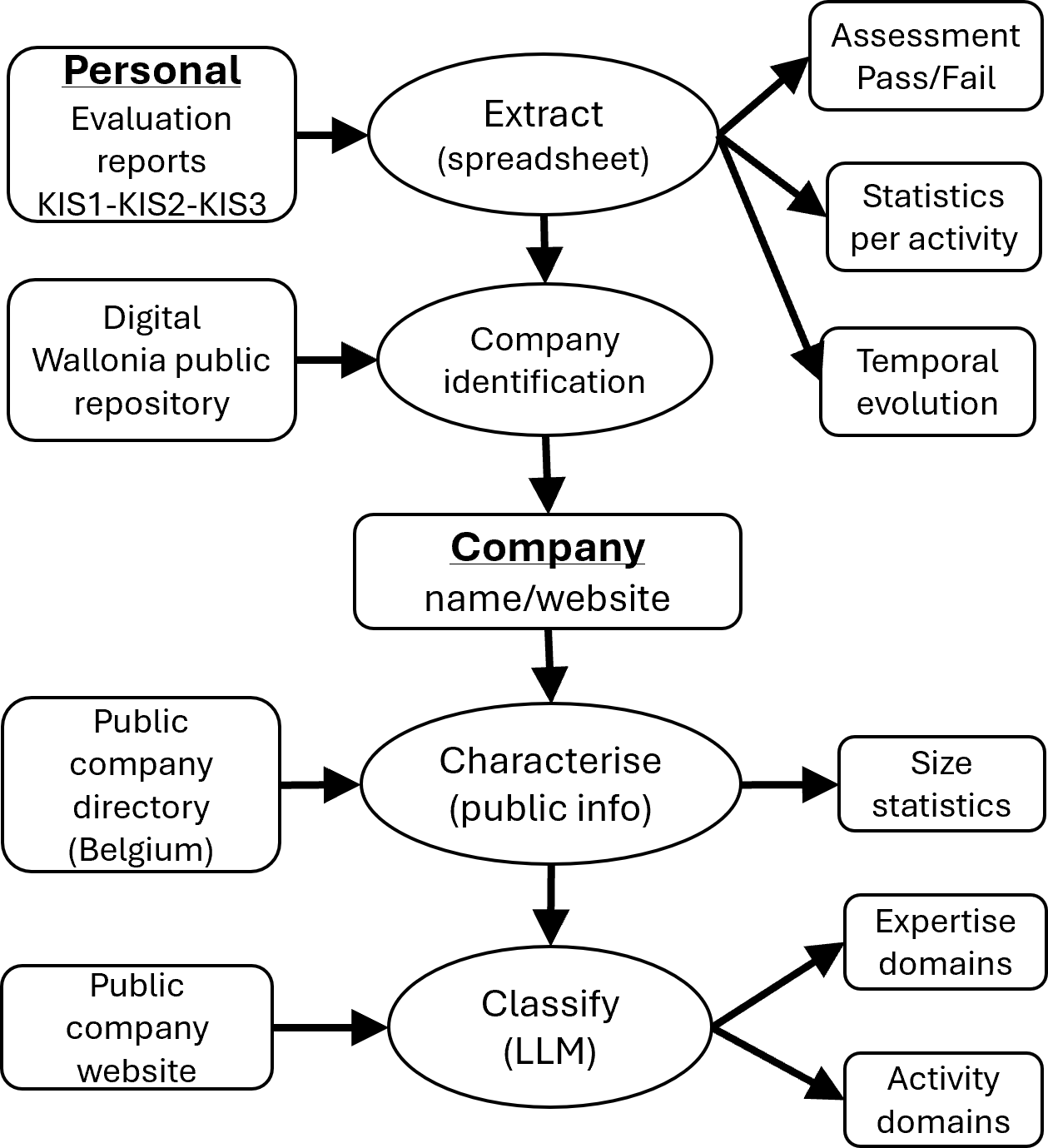}
\caption{Data analysis process}
\label{fig:process}
\end{figure}

The data analysis process is depicted at Figure \ref{fig:process}. The central material is the filled assessment questionnaire described previously. These questionnaires are available in Excel format and are automatically analysed using a Python script taking into account the variations introduced by the evolution of the answer scoring method. Based on extracted company identity, a specific agent is used to retrieve key information from public sources. On one hand, company statistics (e.g., size and age from online directories) are retrieved in a structured manner. On the other hand, information about a company’s business and activity domains is extracted in an unstructured way through website analysis using a LLM, which also provides traceable justification. Specifically, the LLM generates explanations citing website excerpts used to infer particular areas of activity and domain expertise, allowing a human analyst to verify the correctness of the interpretation during the data review phase.

\subsection{Expert and company characteristics}

Over the six-year operation period, we audited 122 experts coming from 87 different companies. Table \ref{tab:dataset} details the breakdown per phase with a first phase somewhat larger than the next ones.

\begin{table}[h]
\centering
\small
\begin{tabular}{|l|c|c|}
\hline
Phase  &  \# experts audited & \# companies \\
\hline
\hline
KIS1 (2019-2021) & 52 & 38 \\
\hline
KIS2 (2022-2023) & 34 & 29 \\
\hline
KIS3 (2024-) & 37 & 30 \\
\hline
Total & 122 & 87 \\
\hline
\end{tabular}
\caption{KIS data set}
\label{tab:dataset}
\end{table}

Table \ref{tab:size} shows the majority of companies are very small, reflecting the Walloon landscape, with only a few medium and large-sized firms. In the smallest companies, the business areas are closely aligned with the (single) expert’s domain of expertise. Between 10\% and 33\% of companies have more than one expert, with a higher proportion in the mid-range. Small companies tend to have a more specialized focus on cybersecurity, whereas larger companies maintain broader IT departments capable of hosting multiple certified cybersecurity experts.

\begin{table}[h]
\centering
\small
\begin{tabular}{|l|c|}
\hline
Size  &  \# companies \\
\hline
\hline
very small (1-10) & 63 \\
\hline
small (11-50) & 12 \\
\hline
medium (51-250) & 6 \\
\hline
big ($>250$) & 4 \\
\hline
not classified & 2 \\
\hline
Total & 87 \\
\hline
\end{tabular}
\caption{Distribution of company size}
\label{tab:size}
\end{table}

\begin{figure}[h]
\centering
\includegraphics[width=0.8\columnwidth]{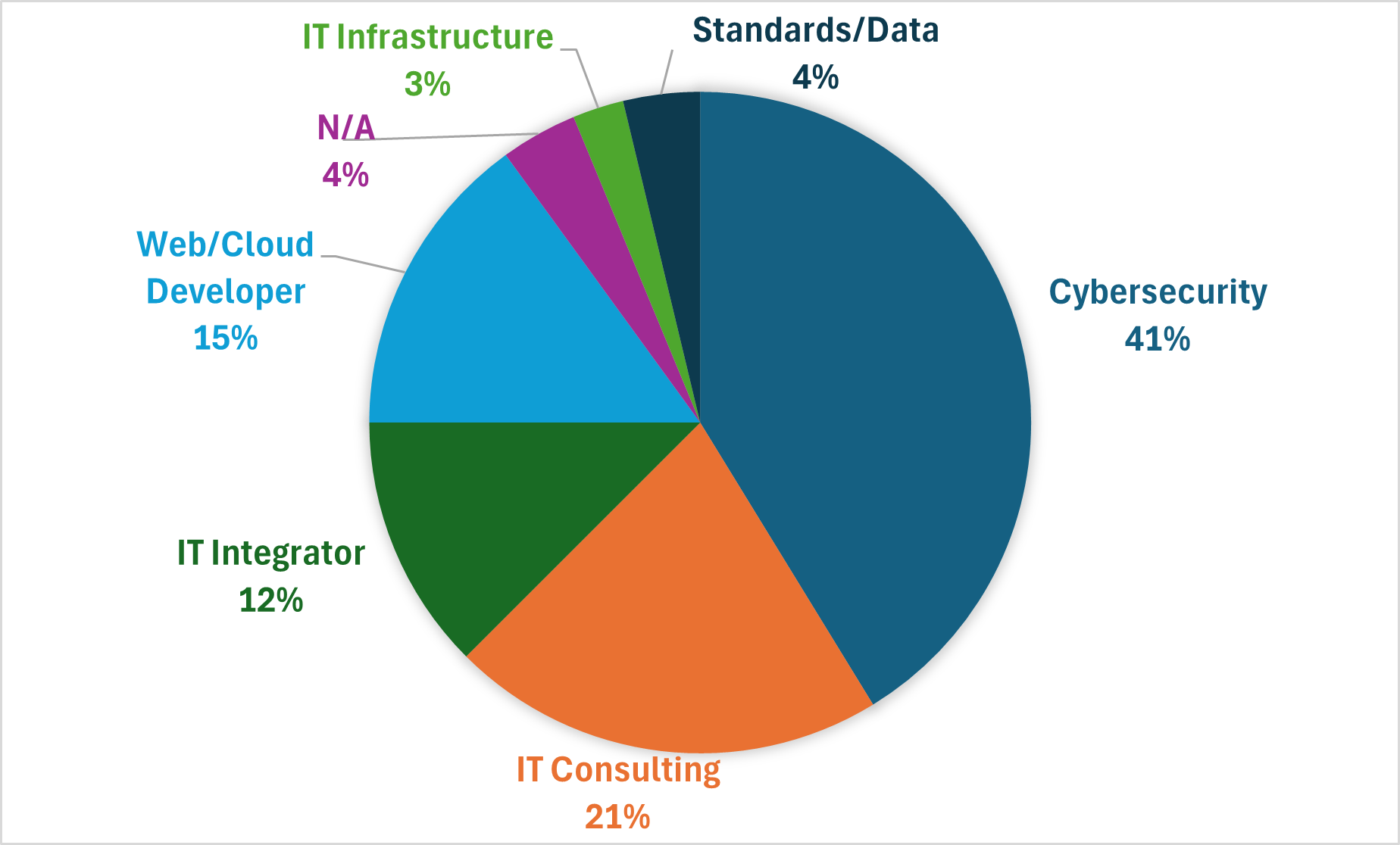}
\caption{Repartition across sectors}
\label{fig:repartition}
\end{figure}

Figure \ref{fig:repartition} presents the distribution across key domains of activities. The majority of actors are focused on cybersecurity. Next come IT consultants, developers, and integrators. Those categories provide a relevant basis which can lead to acceptable KIS experts if they have proven enough cybersecurity knowledge. A smaller portion of companies specialise in infrastructure or in standards and regulations, such as GDPR which is usually not accepted, the latter being out of scope.

\begin{figure}[h]
\centering
\includegraphics[width=0.85\columnwidth]{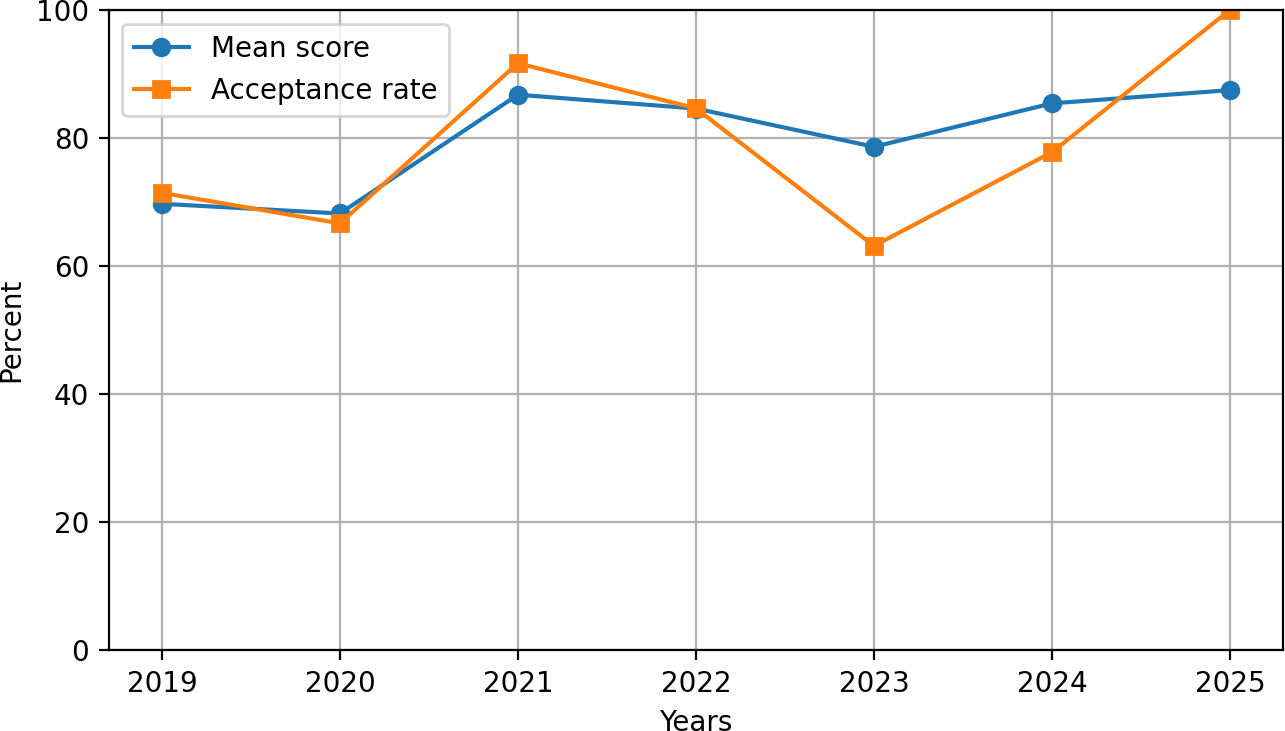}
\caption{Evolution of acceptance rate and scores}
\label{fig:acceptance-score}
\end{figure}

Figure \ref{fig:acceptance-score} shows that in the first years many candidates were lacking cybersecurity expertise, actually too technical or not sufficiently SME-oriented. There were also some dropouts before evaluation. In 2021, after this first wave, a clear improvement emerged, due to better understanding of the KIS requirements and possibly a COVID-related effect with more time to get ready. Post-COVID, the average score remained good, but refusals increased, highlighting a gap between top experts and less skilled ones. Half of refusals resulted from insufficient framework knowledge, while the other half resulted from a lack of audit autonomy. Performance improved again after 2024 with 2025 data being less representative as only 7 audits were conducted before a moratorium still in effect.

\subsection{Phase-level Analysis}

The phase level analysis is based on the five NIST CSF1.1 phases. Table \ref{tab:scores} provides the mean scores across those categories and the 3 main KIS phases. Overall, scores are highest for protection, followed by identification. The later phases relating to Detect, Respond and Recover are consistently weaker. This observation was also reported in a recent literature survey reporting that ``most activities are narrowly focused on the Identify and Protect functions of the NIST CSF with very little work on the other existing functions'' \cite{Chidukwani22}. Despite this, there is also general a tendency toward improvement across all CSF activities.

\begin{table}[h]
\centering
\small
\begin{tabular}{|p{1.1cm}|P{0.6cm}|P{0.7cm}|P{0.7cm}|P{1.0cm}|}
\hline
Phase & KIS1 & KIS2 & KIS3 & Global \\
\hline
\hline
Identify & 74\% & 73\% & 84\% & 76\%  \\
\hline
Protect & 81\% & 77\% & 85\% & 81\%  \\
\hline
Detect & 63\% & 67\% & 77\% & 68\% \\
\hline
Respond	& 61\% & 64\% & 75\% & 66\% \\
\hline
Recover	& 61\% & 69\% & 80\% & 69\% \\
\hline
\end{tabular}
\caption{Evolution of score across KIS phases}
\label{tab:scores}
\end{table}

To better assess the maturity, the check on spontaneous answer is very relevant because it reveals the core knowledge of a candidate and points that are unlikely to be overlooked during an audit. Table  \ref{tab:maturity} shows figures correlated with, and lower than, the total scores as expected. It also confirms the lower performance in Detect, Respond and Recover phases. The evolution trend is also different: it is rather stable for Identify and Protect while it seems to degrade for Detect, Respond and Recover. This points to the need for more specific actions to better support those phases. 

\begin{table}[h]
\centering
\small
\begin{tabular}{|p{1.1cm}|P{0.6cm}|P{0.7cm}|P{0.7cm}|P{1.0cm}|}
\hline
Phase & KIS1 & KIS2 & KIS3 & Global \\
\hline
\hline
Identify & 61\% & 64\% & 67\% & 64\%  \\
\hline
Protect & 72\% & 73\% & 72\% & 72\%  \\
\hline
Detect & 50\% & 53\% & 47\% & 50\% \\
\hline
Respond	& 50\% & 42\% & 35\% & 43\% \\
\hline
Recover	& 50\% & 55\% & 46\% & 50\% \\
\hline
\end{tabular}
\caption{Evolution of spontaneous answer across phases}
\label{tab:maturity}
\end{table}



\subsection{Evolution of the KIS ecosystem}

\begin{figure}[h]
\centering
\includegraphics[width=0.99\columnwidth]{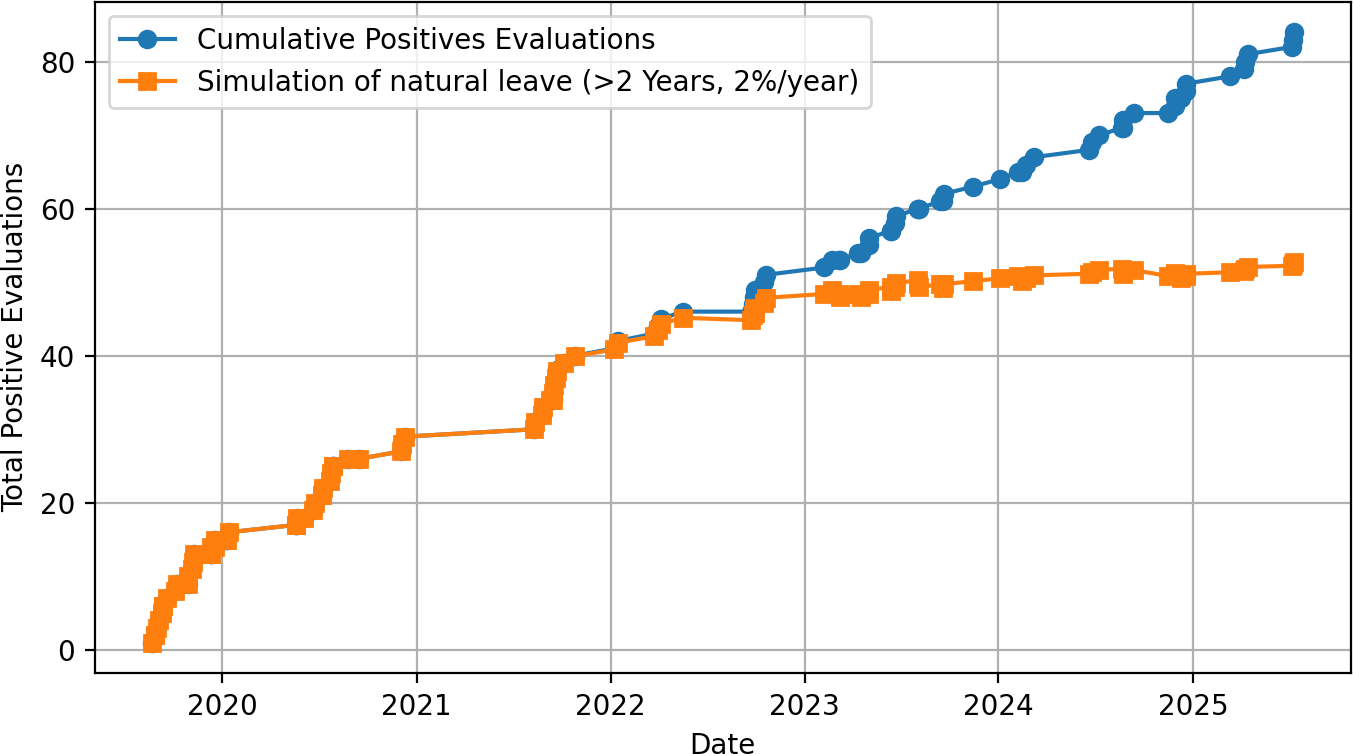}
\caption{Evolution of the KIS ecosystem}
\label{fig:evolution}
\end{figure}

Finally, Figure \ref{fig:evolution} shows the evolution of the ecosystem of labelled experts. The cumulative total shows an intense initial phase, followed by a COVID-related pause and a resumption in summer 2020 with remote interviews, which are still in use. By mid-2021, the transition from KIS2 led to a peak in activity in autumn 2021. Afterwards, activity remained fairly steady, with the KIS3 transition in 2024 occurring without any notable interruption. No new cases have been added since July 2026 due to the moratorium.

The actual size of the ecosystem is however lower with currently 56 active experts. We don't have the evolution details of this actual size over time but we could simulate it using the realistic assumption that most experts remain active for two years, and then a departure rate of 2\% is considered. This allowed us to simulate the lower curve resulting in the current ecosystem. It shows the size remains quite stable under the condition we continue to accept new experts, which is currently suspended.

\subsection{Threats to validity}

This longitudinal study was conducted on our regional dataset with a methodology detailed in Section 3. It may be affected by some validity issues such as:
\begin{itemize}
\item \textit{Internal validity}: there might be bias regarding the experts present in the KIS network vs all experts providing services. However, the fact that the label gives access to a funding scheme for the SME is a strong incentive to join the KIS network. To our knowledge, there are few experts currently operating outside of the network. Our count can be considered representative of the expert community.
\item \textit{External validity}: at this point, we did not attempt to generalise to similar ecosystems in other countries mainly due to the variety of SME support mechanisms with different scopes and formalisation, e.g. CyberEssential/CyberEssential+ in the UK \cite{cyberessentials} close to our Keep IT Secure/CyFun or ExpertCyber label in France \cite{ExpertCyber}. We conducted an initial review of the European landscape of existing initiatives directed towards SMEs \cite{Ponsard18}, but it requires a substantial update to allow a current comparison, combined with the limited availability of public data. This is considered as part of future work by us or by the research community.
\item \textit{Construct validity}: our study only relies on our assessment questionnaire. However, it is based on the sound NIST CSF framework and confirmed with the approach proposed by CyFun at our federal level. Despite this, the full range of skills is not explicitly assessed and is part of the identified improvements.  Note that soft skills (e.g. ability to conduct interviews, to understand the SME context) are part of the current evaluation but were not reported here.
\item \textit{Conclusion validity}: we have collected a substantial dataset which enables sound statistics. The scoring was performed by a pool of two evaluators with one of the experts present for the whole duration. The scoring method was improved but kept a common ground over the whole study.
\end{itemize}

\section{\uppercase{Lessons Learned and Evolution}}

\textbf{The introduction of a cybersecurity labelling } in an initially unregulated environment with self-proclaimed cybersecurity experts could have triggered some concerns. However, most candidates without a minimal track record did not enter the process and this was welcomed both by the experts able to get the label and the SMEs looking after cybersecurity advice. We were also positively surprised by the maturity of some candidates, who had developed well-structured and documented methodologies, including awareness-raising activities within client organisations. This approach has remained unique until now in the way other kind of help are provided to our SMEs but it resulted in confirmed abuse by self-proclaimed experts and in misuse of public money. As a result, a skill validation approach is now being considered for a wider set of support measures with a systematic structuring around well-identified skills.

\textbf{The range of expert skills} considered by KIS is currently not specialised: it mainly consists of the ability to conduct an audit with an SME and the knowledge of basic cybersecurity controls borrowed from reference frameworks such as NIST CSF and CIS. An on-going process is to consider a wider range of skills considering underlying technologies (e.g. web/cloud, IoT, infrastructure,...), norms (GDPR, NIS, CRA) and new relevant fields such as AI or CTI. Different actors have proposed or are working on the definition of skill frameworks such as the European Cybersecurity Skill Framework by \cite{ENISA22}, the French skills matrix \cite{Cybermalveillance23}, the Portuguese competencies framework \cite{CNCS24}. An interesting and still on-going work is to investigate how to best define such skills and how to map them with sector-specific needs e.g. for companies operating in IT, industrial control, health, finance,...


\textbf{Specific evolution needs} were also collected during the interview process. Although the assessment grid was designed for SME maturity and validated prior to deployment, some criteria proved too advanced (e.g., forensic analysis or direct cooperation with local CERTs), while others required finer granularity, such as aligning security policies with the company’s purpose. These adjustments were reviewed between phases with an advisory board including academic experts and KIS practitioners. Beyond refining the evaluation criteria, these exchanges help define SME maturity improvement paths, support awareness-raising for new SMEs, and facilitate the sharing of good practices among experts. As we are now looking to capture more explicitly cybersecurity skills in our process, we need a deeper validation of the skill matrix under design by organizing specific workshops with a pool of KIS labelled experts and the cybersecurity specialists of our governance committee.

\textbf{An identified current weakness} is the focus on Govern, Identify and Protect steps while Detect, Respond and Recover are lagging behind. This was also identified in a literature survey reporting that cyber security incident detection, response and recovery are hardly accounted for in research work for SMEs and stressed the need to enhance cyber resilience \cite{Chidukwani22}. A practical instrument to increase skills is to give access to cyber ranges with typical SME attack scenarios so experts can learn measures that are useful at those later stages and be able to make better advice in their consultancy work. Such an initiative is currently being deployed locally.

\textbf{The role of the expert ecosystem} is crucial for deploying cybersecurity. Literature highlights that IT companies play a cascading role in disseminating cybersecurity best practices to micro and small businesses. Especially, a UK study showed that the primary channel through which SMEs access cybersecurity information is through their local IT or cyber support providers and not from government bodies or awareness campaigns \cite{Cartwright23}. Such an ecosystem should be supported in order to sustain its development. So far, this has been the case with the KIS initiative which has reached a stable size. It needs to keep growing given the ongoing expansion of SME digitalisation increasing exposure to cyber threats and the acceleration of the volume/complexity of attacks especially AI-driven.

\textbf{About the support for regulation.} When KIS went into operation, GDPR was a great incentive for raising awareness about cybersecurity inside SMEs \cite{GDPR}. One inconvenience is that some GDPR consultancy companies also presented themselves as cybersecurity advisors based on their data protection expertise. The fact that regulatory compliance may not be funded helped solve the issue. However, it remains important that regulation requirements which form the security foundation layer are mastered by the cybersecurity experts. This is currently quite the case for data security with GDPR but new regulations and directives have come into force since then, such as Network and Information Systems Directive \cite{NIS2} and soon the Cyber Resilience Act \cite{CRA}.

\section{\uppercase{Conclusion \& Next Steps}}

In this paper, we provided a longitudinal analysis of the cybersecurity experts ecosystem that developed in Belgium (Wallonia) to help secure local SMEs over the period 2019-2025. We explained how the Keep IT Secure initiative was designed and operated to label those experts and we analysed the data systematically collected during interviews over that operation period. We identified key strengths like its stability and alignment with the federal Cyber Fundamentals framework. We also identified some weaknesses such as the lower maturity in Detect, Respond and Recover phases. The process does not impose a specific methodology but follows the expert methodology based on a grid that already ensures a very good alignment with the Belgian CyFun certification scheme. 

At this stage, our work is primarily observational and does not attempt to generalise or compare with similar cybersecurity ecosystems in other countries, given the substantial work required to identify current initiatives and collect data. This would be best achieved at the European level, e.g. by \cite{ENISA} or a collaborative network like \cite{ECSO}.

Our future work will also aim to maintain alignment with NIST CSF 2.0 and CyFun 2025. We are actively working to enhance the reference skills framework by including updates for new regulations and AI-related threats, and to extend our interview process and questionnaire to evaluate these skills effectively. Finally, in a wider context, we are contributing to the elaboration of cyber range training scenarios to improve practical exercises for detecting, responding, and recovering from typical attacks targeting SMEs.



\bibliographystyle{apalike}

{\small
\bibliography{biblio}}

\vfill
\end{document}